\documentclass{vgtc}                          % final (conference style)
\ifpdf%                                % if we use pdflatex
  \pdfoutput=1\relax                   % create PDFs from pdfLaTeX
  \usepackage{graphicx}                % allow us to embed graphics files
  \DeclareGraphicsExtensions{.pdf,.png,.jpg,.jpeg} % for pdflatex we expect .pdf, .png, or .jpg files
\else%                                 % else we use pure latex
  \usepackage{graphicx}                % allow us to embed graphics files
  \DeclareGraphicsExtensions{.eps}     % for pure latex we expect eps files
\fi%

\graphicspath{{figures/}{pictures/}{images/}{./}} % where to search for the images

\usepackage{microtype}                 % use micro-typography (slightly more compact, better to read)
\PassOptionsToPackage{warn}{textcomp}  % to address font issues with \textrightarrow
\usepackage{textcomp}                  % use better special symbols
\usepackage{mathptmx}                  % use matching math font
\usepackage{times}                     % we use Times as the main font
\usepackage{cite}                      % needed to automatically sort the references
\usepackage{tabu}                      % only used for the table example
\usepackage{booktabs}                  % only used for the table example
\usepackage{comment} 
\usepackage{balance}
\onlineid{0}

\vgtccategory{Research}

\vgtcinsertpkg

\title{Towards Universal Interaction for Extended Reality}

\author{Pascal Knierim\thanks{e-mail: pascal.knierim@uibk.ac.at}\\ %
        \scriptsize University of Innsbruck, Austria %
\and Thomas Kosch\thanks{e-mail: thomas.kosch@hu-berlin.de}\\ %
        \scriptsize Humboldt University of Berlin, Germany %
        }

\abstract{
Extended Reality (XR) is a rapidly growing field offering unique immersive experiences, social networking, learning, and collaboration opportunities. The continuous advancements in XR technology and industry efforts are gradually moving this technology toward end consumers. However, a universal one-size-fits-all solution for seamless XR interaction still needs to be discovered. Currently, we face a diverse landscape of interaction modalities that depend on the environment, user preferences, task, and device capabilities. Commercially available input methods like handheld controllers, hand gestures, voice commands, and combinations of those need universal flexibility and expressiveness. Additionally, hybrid user interfaces, such as  smartwatches and smartphones as ubiquitous input and output devices, expand this interaction design space. In this position paper, we discuss the idea of a universal interaction concept for XR. We present challenges and opportunities for implementing hybrid user interfaces, emphasizing \textit{Environment}, \textit{Task}, and \textit{User}. We explore the potential to enhance user experiences, interaction capabilities, and the development of seamless and efficient XR interaction methods. We examine challenges and aim to stimulate a discussion on the design of generic, universal interfaces for XR.

} % end of abstract

\CCScatlist{ 
  \CCScat{H.5.1}{Information Interfaces and Presentation}%
{Multimedia Information Systems}{Artificial, augmented, and virtual realities};
  \CCScat{H.5.2}{Information Interfaces and Presentation}%
{User Interfaces}{Input devices and strategies};
}

\keywords{Extended Reality, Hybrid User Interfaces, Interaction}

\begin{document}

\maketitle

\section{Introduction and Background}
Extended reality (XR) is a rapidly growing field offering novel opportunities to augment realities, explore virtual environments, engage in social networks, improve learning, and collaborate with peers~\cite{10.1145/3544548.3581072, 10.1145/3491102.3517593}. Continuous research advances and efforts by industries slowly move this technology toward end-consumers while, at the same pace, a diverse landscape of interaction modalities is growing. Despite these extensive efforts, a universal, adaptable solution for seamless interaction in XR has yet to be discovered. This is because the most suitable interaction modality can differ based on various factors, such as the user's preferences, the environment, the interaction scenario, and the device's capabilities. Currently, commercially available XR devices offer different input methods ranging from dedicated handheld controllers, mid-air hand gestures, and voice commands to a mix of hand or gaze gestures. There have also been recent advancements in research that extend the scope of input interaction, for example, by utilizing wrist-worn sensors to detect touch events in the physical space~\cite{9417793} or using smartphones as versatile input devices~\cite{KnierimHeinSchmidtKosch+2021+49+61}.

While hybrid interfaces utilizing smartphones or smartwatches can provide superb usability, better user experiences, and lower workloads~\cite{doi:10.1080/10447318.2022.2160537}, these new modalities also introduce new challenges, such as the need to integrate multiple devices and to develop and learn new interaction techniques. While interaction modalities are improving, no interaction modalities suit every use case. We envision hybrid interfaces to minimize the gap between the intended interaction in XR and the available input device.

Mouses, keyboards, and touchpads have become integral to today's mobile and stationary computing and serve as generic, high-throughput input devices. Similarly, smartphones support direct gestures and touch input, providing continuous haptic and visual feedback (e.g., through panning or zooming). Such metaphors are still missing in the XR interaction space. Hybrid user interfaces address the lacking flexibility of XR experiences and further broaden the design space. The advantages of these different technologies combined create unique interaction concepts and ultimately could lead to a universal interaction concept of XR.

\begin{figure}
    \centering
    \includegraphics[width=\linewidth]{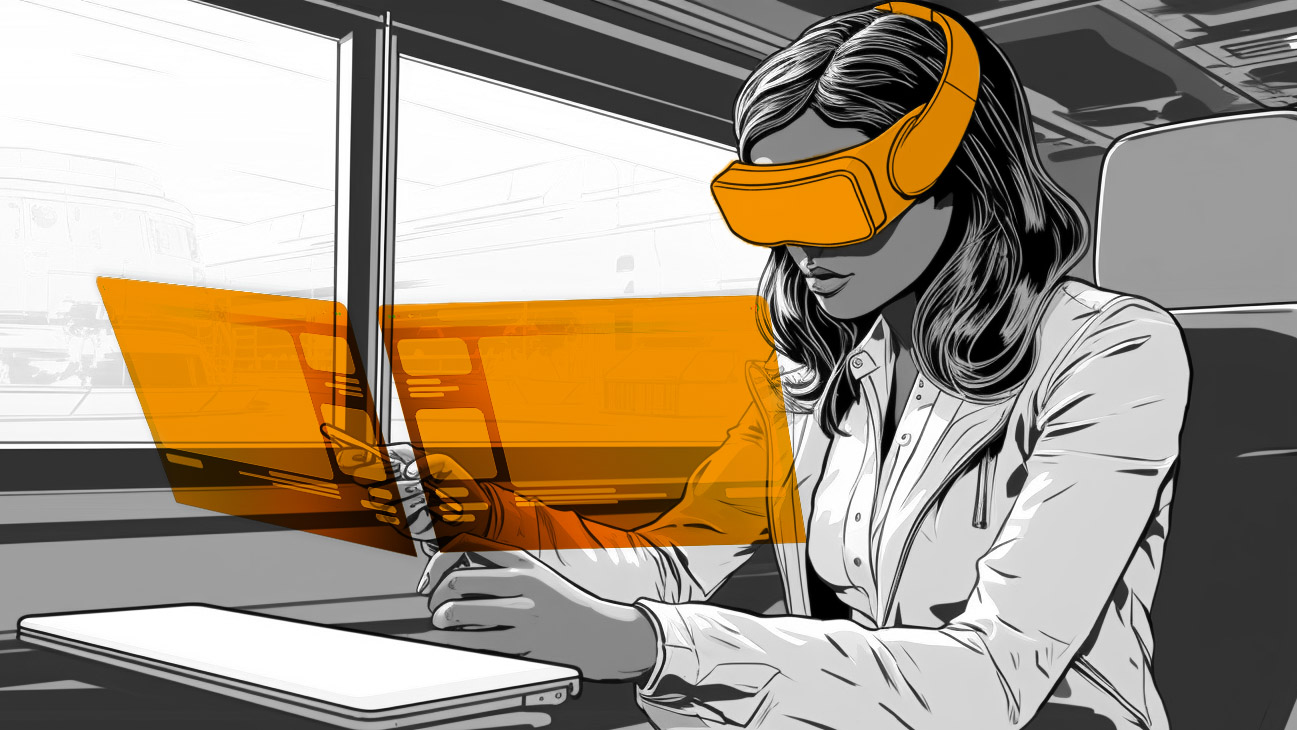}
    \caption{Hybrid interfaces pave the way towards universal interaction, enabling \textit{users} to accomplish various \textit{tasks} in varying \textit{environments} intuitively. For instance, while commuting on a train and browsing the web. Today, hybrid interfaces allow exploring interaction concepts in these challenging conditions.}
    \label{fig:enter-label}
\end{figure}

The idea of interacting in XR beyond using freehand gestures, a typical implementation for XR interaction, is not novel. However, prolonged use of freehand gestures leads to discomfort and requires additional space~\cite{10.1145/3544548.3581461}. Hence, researchers investigate alternative XR interaction modalities~\cite{10.1145/2556288.2557130, 10.1145/3025453.3025523}. For typing tasks, a typical action for knowledge workers, a hologram of a keyboard is displayed, and mid-air touch gestures are used to select individual characters. Consequently, typing can become a tiresome activity.

Past research investigated how different representations of keyboards and hands influence typing efficiency and experience. Grubert et al.\cite{8446250} showed that well-designed XR experiences can significantly lower text entry error rates. Likewise, we~\cite{10.1145/3173574.3173919} demonstrated that high typing performance could be achieved by optimizing the visualization of hands in VR, especially for inexperienced typists. As an alternative to static offices, nomadic offices, where working environments are visualized in XR, are increasingly gaining attention (cf. Fig.~\ref{fig:enter-label}). To foster a productive and meaningful working environment, currently, nomadic offices rely on well-known keyboard input \cite{9603999}. While falling behind on copy editing, nomadic offices with hybrid interfaces can provide ample display space and enhanced privacy~\cite{10.1145/3334480.3382920}. Hubenschmid et al. confirmed that augmenting a smartphone with a desktop monitor-sized extension represents the optimal design~\cite{10.1145/3544548.3581438}. In addition, we showed that a smartphone-based AR controller results in significantly faster and more accurate object manipulation with reduced task load than state-of-art mid-air gestures~\cite{KnierimHeinSchmidtKosch+2021+49+61}. While these types of hybrid interfaces can increase the quality of interaction and further support users in their tasks, unique challenges exist in designing such environment~\cite{ZagermannHubenschmidBalestrucciFeuchtnerMayerErnstSchmidtReiterer+2022+145+154}.

\begin{figure}
    \centering
    \includegraphics[width=.9\linewidth]{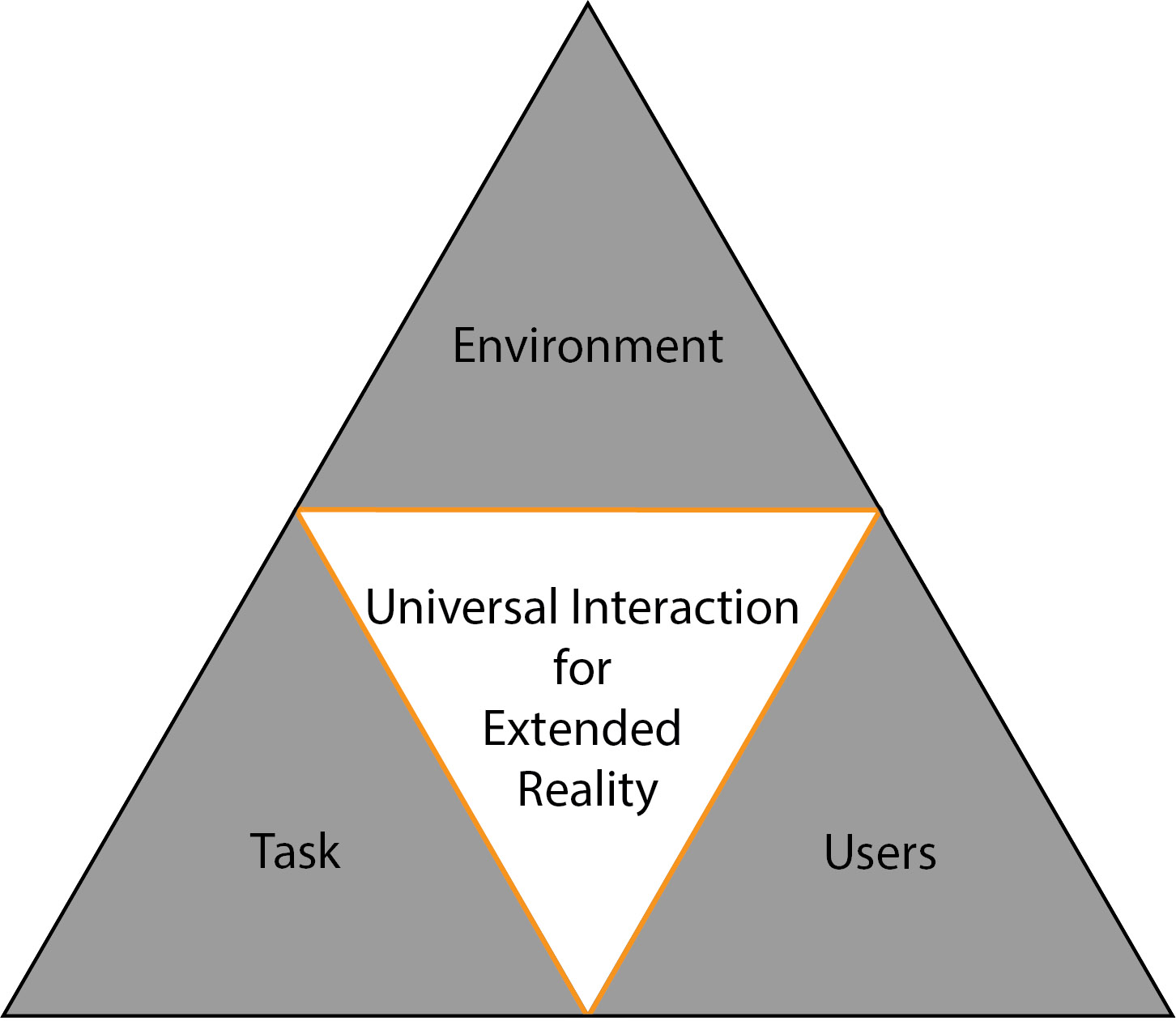}
    \caption{Pillars contributing to universal interaction for XR. \textbf{Environment:} The user's physical context influences XR interactions, and understanding this can enable seamless digital integration, but mobile scenarios and privacy concerns may pose challenges. \textbf{Task:} XR applications offer varied experiences requiring different interaction levels, from immersive, prolonged gameplay to simple notification dismissals. \textbf{User:} A universal XR interaction concept must consider user diversity and strive for flexibility, ease of use, and accessibility, despite current technological limitations.}
    \label{fig:enter-label_01}
\end{figure}

Here, we stress the development of a universal interaction concept for XR, comparable to the gold standards of multitouch for smartphones or mouse and keyboard for desktop computing. We are confident that hybrid interfaces allow us to study potential solutions now and ultimately bring the same utility, like keyboard and mouse or touchscreens, to the domain of XR. By discussing universal interaction today, we assert that such principles can enable the research community to devise usable XR interfaces that are inherently intuitive and effortless. We aim to discuss and lay down the foundations for exploring how universal interaction, for XR, can foster more engaging and intuitive user experiences. We strive to investigate the existing challenges and opportunities, thereby underlining the potential to augment user experience.

\section{Universal Interaction for Extended Reality}
Hybrid user interfaces for XR represent an approach to complement XR interfaces, seamlessly combining different devices with diverse input and output modalities to offer users a rich, intuitive, and more immersive experience. We argue that the resulting considerable interaction design space is based on three pillars (cf. Fig~\ref{fig:enter-label_01}): \textit{Environment}, \textit{Task}, and \textit{User}. To achieve a universal interaction concept for XR, the input and output characterization needs to accomplish the resulting needs in the following three areas bridging the current gap between custom-tailored and intuitive interaction.

\subsubsection*{Environment}
The user's physical environment and context during an XR experience fundamentally impact the potential interactions. Understanding the user's surroundings allows for seamless integration of digital content within the real world and facilitates improved interaction with different interfaces. For example, mobile scenarios might be limited in space for interaction, and additional resources may also be constrained. Furthermore, various scenarios present distinct privacy and security concerns, such as private, semi-private, or public environments, including private car rides, plane travel, or train commuting. Thus, considering the different and varying aspects of the environment while designing interaction concepts requires the implication of flexibility while maintaining a consistent concept.  

%Hence, integrating suboptimal with the pillar \textit{Environment}. 
%Consequently, the pillars \textit{Task} or \textit{User} might be fulfilled when a mobile scenario that suits the user's requirements is necessary.

\subsubsection*{Task}
As previously outlined, XR applications offer diverse experiences, encompassing entertainment, social networking, learning, and collaborative work. Due to this manifold nature, these applications involve different tasks, resulting in varying interaction timespans and required input. For instance, in an immersive game, users may want to interact over a prolonged time with digital artifacts and navigate intuitively through the virtual environment. Yet, there are also instances where interaction is minimal, such as a simple confirmation to dismiss a message or notification. Hybrid UIs can help by providing task-specific adaptations while keeping the consistency across different domains high. Ultimately, this would lead to familiarity with certain interactions, enhancing efficiency and simplicity. 

\subsubsection*{User}
Users have diverse backgrounds, abilities, and familiarity with digital products. A universal interaction concept for XR must acknowledge these variabilities by providing flexibility in use, ensuring low physical effort during extended input periods, or providing the possibility for shortcuts. Currently, specific XR hardware and software technologies are not fully accessible. However, future research can address this shortcoming and make XR applications more inclusive and usable for everyone.

By following the idea of a universal interaction concept for XR, developers and designers can create user interfaces that are versatile, user-friendly, intuitive, and inclusive. Already today, hybrid user interfaces offer the potential to bridge the gap between existing tailor-made input and output concepts and bring us one step closer to natural XR interaction. By offering multi-modal interaction, adaptive interfaces, and contextual integration, hybrid UIs can enhance usability and reduce cognitive load~\cite{10.1145/3205873.3210708}.

\section{Future Directions of Universal Interaction}
Here, we point out research directions that tackle the current challenges of universal interaction.  

\subsection*{Challenge 1: Intuitive Interaction in Universal Spaces} The first and most significant challenge lies in creating an intuitive and versatile interaction concept. To effectively reach a diverse user base, these concepts should be self-explanatory and easy to grasp regardless of users' technical familiarity or background. Further, they need to be adjustable to serve a multitude of tasks. 

\subsection*{Challenge 2: Opportune Utilization of Attention} Attention is a scarce resource that decreases over a day. Frequent attention shifts, for example, created by smartphone notifications, lead to faster depletion of attentional resources. Here, the attention shifts characteristic of XR environments pose another challenge; as users navigate between hybrid interfaces, the physical and digital realms, interfaces must be designed to manage and facilitate this transition smoothly. 

\subsection*{Challenge 3: Technical Boundaries} Power consumption and battery life remain critical technical challenges in mobile scenarios, as the computational demands of immersive XR applications can drain resources quickly. Solutions may involve more energy-efficient algorithms or hardware improvements. A potential clutter of hybrid input devices, from hand-held controllers to haptic suits, complicates the user experience and calls for streamlined, perhaps even embodied solutions. 

\subsection*{Challenge 4: Fragmentation} Lastly, the risk of fragmented ecosystems across different head-mounted display manufacturers could hinder universal interaction, accessibility, and usability. Therefore, industry-wide standards or interoperability protocols might be a critical area of focus to ensure a consistent user experience across platforms. As we navigate these challenges, the future of XR will likely be shaped by a commitment to inclusivity, user-centered design, and cross-platform consistency.

\section{Conclusion}
Integrating hybrid user interfaces in XR introduces several challenges, opportunities, and future research directions. In this position paper, we outline the potential of hybrid interfaces in universal extended reality (XR) interaction scenarios. First, we outline three aspects that need to be considered when designing a universal interaction concept for XR, namely the \textit{Environment}, \textit{task}, and \textit{user}. Further, we identify four essential challenges based on our past experiences and related work: \textit{Intuitive Interaction in Universal Spaces},  \textit{Opportune Utilization of Attention}, \textit{Technical Boundaries}, and \textit{Fragmentation}. Inconsistent interaction paradigms, fragmented ecosystems, limited battery life, excessive device clutter, the need for intuitive learning of new interaction methods, and the necessity to manage shifting attention are all key aspects that need careful consideration. By recognizing and actively tackling these issues in our research community, designers and developers have the opportunity to craft user interfaces that provide smooth, intuitive, and captivating XR experiences. We are confident that our position paper will spark discussions about hybrid interfaces, setting the course for feasible universal interaction in XR.

% In conclusion, the implementation of hybrid user interfaces in XR presents several challenges. Overcoming the fragmented nature of the ecosystem, addressing battery life limitations, managing device clutter, facilitating the learning of new interaction concepts, and handling attention shifts are critical areas that require careful consideration. By acknowledging and proactively addressing these challenges, designers and developers can create user interfaces that offer seamless, intuitive, and engaging XR experiences. This approach holds the potential to establish a new universal gold standard for XR interaction.

%\acknowledgments{
%The authors wish to thank A, B, C. This work was supported in part by
%a grant from XYZ.}
\balance
\bibliographystyle{abbrv-doi}

\bibliography{main}
\end{document}